\documentclass[12pt]{iopart}
\usepackage{graphicx}
\usepackage{subfigure}
\usepackage{caption2}
\usepackage{float}
\usepackage{cite}

\begin{document}

\title[Modulational Stability and Dark Solitons in $\chi^{(2)}$ Nonlin.
Waveguide with Grating]{Three-Wave Modulational Stability and Dark Solitons
in a Quadratic Nonlinear Waveguide with Grating}

\author{Arthur Gubeskys and Boris A. Malomed}

\address{Department of Interdisciplinary Studies, School of Electrical
Engineering, Faculty of Engineering, Tel Aviv University, Tel Aviv 69978,
Israel}

\begin{abstract}
We consider continuous-wave (CW) states and dark solitons (DSs) in a system
of two fundamental-frequency (FF) and one second-harmonic (SH) waves in a
planar waveguide with the quadratic [$\chi ^{(2)}$] nonlinearity, the FF
components being linearly coupled by resonant reflections on the Bragg
grating (the same model is known to support a great variety of bright
solitons). We demonstrate that, in contrast with the usual situation in
spatial-domain $\chi ^{(2)}$ models, CW states with the phase shift $\pi /2$
between the FF and SH components are modulationally \emph{stable} in a broad
parameter region in this system, provided that the CW wavenumber does not
belong to the system's spectral gap. Stationary fundamental DSs are found
numerically, and are also constructed by means of a specially devised
analytical approximation. Bound states of two and three DSs are found too.
The fundamental DSs and two-solitons bound states are \emph{stable} in all
the cases when the CW background is stable, which is shown by dint of
calculation of the corresponding eigenvalues, and verified in direct
simulations. Tilted DSs are found too. They attain a maximum contrast at a
finite value of the tilt, that does not depend on the $\chi ^{(2)}$ phase
mismatch. At a maximum value of the tilt, which\ grows with the mismatch,
the DS merges into the CW background. Interactions between the tilted\
solitons are shown to be completely elastic.
\end{abstract}

\pacs{42.65.Tg, 05.45.Yv}

\submitto{\JOA}

\eads{\mailto{gubeskys@post.tau.ac.il}, \mailto{malomed@post.tau.ac.il}}

\maketitle

\section{Introduction}

Optical dispersive and diffractive media with quadratic [$\chi
^{(2)}$] nonlinearity are well known for their potential to
support various types of solitons, which have been a subject of
intensive studies \cite{Jena,Buryak}. In most cases, $\chi ^{(2)}$
solitons are experimentally observed in the spatial domain, as the
small size of available samples makes it difficult to accommodate
the dispersion length of temporal solitons. Nevertheless, using
special techniques, such as tilted wave fronts, it is possible to
induce a strong artificial dispersion and thus create temporal
solitons in available $\chi ^{(2)}$ optical crystals
\cite{PaoloTemporal}.

Another challenge for the experiment was creation of $\chi ^{(2)}$
dark solitons (DSs) and their two-dimensional counterparts
(optical vortices). For the first time, DSs in $\chi ^{(2)}$
models were considered in the cascading limit, which corresponds
to a large phase mismatch between the fundamental-frequency (FF)
and second-harmonic (SH) waves \cite{Werner94}. In this limit, the
$\chi ^{(2)}$ model reduces to the nonlinear Schr\"{o}dinger (NLS)
equation with the Kerr [$\chi ^{(3)}$] nonlinearity. Beyond the
cascading limit, a single analytical solution \cite{Hayata94}, and
a continuous family of numerical solutions for fundamental and
twin-hole DSs \cite{Buryak95b, Buryak95c} were found. All these
solitons are unstable in the spatial domain, due to the
modulational instability (MI) of the CW (continuous-wave)
plane-wave background which supports the DS. The only possibility
to avoid the MI was found in the temporal domain, in the case when
the group-velocity-dispersion coefficients have opposite signs at
the FF and SH \cite{Trillo95, Kennedy96, He96} (in fact, such a
case is quite realistic in the temporal domain \cite{Kale}).
Nevertheless, effectively stable spatial-domain vortices,
supported by a background of finite extension, were experimentally
created in the case of negative phase mismatch between the FF and
SH waves \cite{PaoloVortex}. In the latter case, the MI had the
character of a convective instability, so that perturbations were
carried away from the vortex's core faster than they grew due to
the MI.

Additional possibilities for creation of solitons are offered by a
combination of $\chi ^{(2)}$ nonlinearity with Bragg gratings
(BGs). The grating may be introduced in both the time-domain and
spatial-domain models, but in the latter one it is a more
realistic feature, amounting to a system of parallel
\textquotedblleft scratches\textquotedblright\ drawn on the
surface of a quadratically nonlinear planar waveguide. In that
case, the generic $\chi ^{(2)}$ model is a four-wave one, as both
the FF and SH components include two counter-propagating waves
(detailed descriptions of the model can be found in the reviews
\cite{Jena} and \cite{Buryak}). The MI of CW solutions was studied
in the four-wave model, with a conclusion that they are unstable
in most cases. Stability regions for the CW states were also
found, close to the transition to the NLS limit, but only in a
rather exotic case when the BG strength at the SH is larger than
at the FF \cite {He99}. A stable dark bi-soliton in this model was
found numerically near the band edge \cite{Conti97a}.

In this paper, our aim is to investigate the MI of CW states, and,
in the case when they are stable, to find DSs in a
\emph{three-wave}\ spatial-domain system, in which two FF
components are linearly coupled through the resonant scattering on
the BG. Simultaneously, the quadratic nonlinearity couples the two
FF components to the SH wave. The propagation direction $z$ is
parallel to the \textquotedblleft scratches" which form the BG.
Then, the above-mentioned couplings can be achieved by choosing FF
carrier wave vectors to be of equal magnitude and making opposite
angles with the $z$ axis, while the SH wave vector is parallel to
it. Thus, only the FF components are affected by the grating,
while the SH wave is subject to diffraction. This model was
introduced in Ref. \cite{Mak98} in the context of bright solitons.
Stable gap solitons, as well as bright \textit{embedded solitons}
(which exist inside the continuous part of the system's linear
spectrum \cite{Champneys2000}) have been found in this three-wave
model.

The paper is structured as follows. The model is formulated in
section \ref {sec:model}, where we also give an estimate for
relevant values of the physical parameters. In section \ref{MI},
we identify two types of CW solutions and analyze their
modulational stability. It is shown that, on the contrary to other
$\chi ^{(2)}$ models in the spatial domain, in the present case
one type of the CW solutions is \emph{stable} in a broad range of
parameters. DSs supported by the stable CW background are
considered in section \ref{dark}. For stationary fundamental
solitons, we find an approximate analytical and direct numerical
solutions. \textit{Tilted solitons} (with a slant relative to the
propagation axis $z$), as well as bound states of two DSs, are
found too. Section \ref{stability} presents results for the
stability and interactions of the DSs. It is shown that both the
fundamental solitons and their bound states are \emph{stable} in
the whole region where the respective CW is modulationally stable.
The work is concluded by section \ref{conclusions}.

\section{The model}

\label{sec:model}

We consider a quadratically nonlinear planar waveguide with a
spatial BG in the form of scores directed parallel to the
propagation direction $z$. As shown in Ref. \cite{Mak98}, spatial
evolution of two components $u_{1}$ and $u_{2}$ of the complex FF
field, and the complex SH field $u_{3}$ obey the following system
of normalized equations:
\begin{equation}
\begin{array}{l}
i\left( u_{1}\right) _{z}+i\left( u_{1}\right) _{x}+u_{2}+u_{3}u_{2}^{\ast
}=0, \\
i\left( u_{2}\right) _{z}-i\left( u_{2}\right) _{x}+u_{1}+u_{3}u_{1}^{\ast
}=0, \\
2i\left( u_{3}\right) _{z}+D\left( u_{3}\right)
_{xx}+u_{1}u_{2}-qu_{3}=0.
\end{array}
\label{evolution}
\end{equation}
Here $z$ and $x$ are, respectively, the propagation and transverse
coordinates, the subscripts stand for partial derivatives, the
asterisk means for the complex conjugation, $q$ is the
phase-mismatch parameter, and $D$ is an effective diffraction
coefficient, defined with regard to the fact that
Bragg-reflectivity and $\chi ^{(2)}$ constants are normalized to
be $1$. We notice that the complex conjugation applied to Eqs.
(\ref{evolution}), and the change of the notation $u_{1}^{\ast
}\rightarrow u_{1},u_{2}^{\ast }\rightarrow -u_{2},u_{3}^{\ast
}\rightarrow u_{3}$, $q\rightarrow -q$ transforms $D$ to $-D$,
therefore we may assume, without loss of generality, that $D$ is
always positive (while $q$ may be positive, negative, or zero).

Stationary solitary-wave solutions are looked for as
\begin{equation}
u_{1,2}=U_{1,2}(x-cz)\exp (ikz),~u_{3}=U_{3}(x-cz)\exp (2ikz),
\label{general}
\end{equation}
where real $k$ and $c$ are the propagation constant and tilt of the
corresponding optical beam in the $\left( x,z\right) $ plane, and complex
functions $U_{1,2,3}$ are to be found from the equations
\begin{equation}
\begin{array}{l}
i(1-c)U_{1}^{\prime }-kU_{1}+U_{2}+U_{3}U_{2}^{\ast }=0, \\
-i(1+c)U_{2}^{\prime }-kU_{2}+U_{1}+U_{3}U_{1}^{\ast }=0, \\
DU_{3}^{\prime \prime }-2icU_{3}^{\prime }-\alpha
U_{3}+U_{1}U_{2}=0,
\end{array}
\label{moving_model}
\end{equation}
with $\alpha =4k+q$ and the prime standing for $d/d(x-cz)$. For symmetric
(untilted) solutions with $c=0$, one may set $U_{1}(x)=-U_{2}^{\ast
}(x)\equiv U(x)$, $U_{3}(x)\equiv V(x)$, where the function $V(x)$ is real,
which yields a simplified system,
\begin{equation}
\begin{array}{l}
iU^{\prime }-kU-U^{\ast }-VU=0, \\
DV^{\prime \prime }-\alpha V-|U|^{2}=0.
\end{array}
\label{model}
\end{equation}

In physical units, the same value of the power density of the CW
background in the LiNbO$_{3}$ planar waveguide that was used in
the experiments with bright spatial solitons \cite{experiment},
i.e., $\sim 20$ W/$\mu $m at the pump wavelength $\simeq 1.3$ $\mu
$m, may be assumed, in combination with the BG reflectivity $\sim
1$ cm$^{-1}$, which is a typical value for weak gratings. Then,
the physical value of the wavenumber detuning corresponding to
$k>1$, i.e., just above the upper edge of the spectral gap, which
is shown below to be appropriate for DSs, is about $\simeq 1$
cm$^{-1}$. The range of the parameters $\alpha $ and $D$ relevant
to the experiment can be determined as in Ref. \cite{Mak98}, where
the normalized parameters of the model were related to physical
ones. In particular, $\alpha $ may take values in a wide range,
$4\,_{\sim }^{<}~\alpha \,_{\sim }^{<}~100$. For example, the
large phase mismatch used for the experimental generation of
vortices in a $\chi ^{(2)}$ media in Ref. \cite{PaoloVortex}
corresponds to $\alpha \simeq 60$. Realistic values of the
effective SH diffraction parameter are $0.1\,_{\sim
}^{<}~D\,_{\sim }^{<}~1$, depending on the angle between the FF
wave vectors and the $z$ axis. The angle must be, however, small
enough for the applicability of the paraxial approximation, which
is implied in Eqs. (\ref{evolution}).

\section{Modulational stability of CW solutions}

\label{MI}

Equations (\ref{model}) give rise to two types of CW solutions. The first of
them is a real one, with in-phase FF and SH components,
\begin{equation}
U=U_{0}\equiv \pm \sqrt{\alpha (k+1)},V=V_{0}\equiv -(k+1),
\label{cw-two-wave-real}
\end{equation}
which exists provided that $\alpha (k+1)\geq 0$. Below, it will be referred
to as a PR (\textit{pure-real}) solution. In the other solution, the phase
of the FF component is shifted by $\pi /2$ against the SH,
\begin{equation}
U=U_{0}\equiv \pm i\sqrt{\alpha (k-1)},V=V_{0}\equiv 1-k.
\label{cw-two-wave-imag}
\end{equation}
This solution, with a \textit{purely imaginary} FF field, will be
accordingly called a PI one (of course, the SH is real in it). The PI
solution exists provided that $\alpha (k-1)\geq 0$.

To investigate the modulational stability of the solutions, we consider a
perturbed one in the form of

\begin{equation}
\begin{array}{l}
u_{1}=[U_{1}^{(0)}+a_{1}^{(0)}\exp (-i\lambda z+i\Omega x)+b_{1}^{(0)}\exp
(i\lambda ^{\ast }z-i\Omega x)]\exp (ikz), \\
u_{2}=[U_{2}^{(0)}+a_{2}^{(0)}\exp (-i\lambda z+i\Omega x)+b_{2}^{(0)}\exp
(i\lambda ^{\ast }z-i\Omega x)]\exp (ikz), \\
u_{3}=[U_{3}^{(0)}+a_{3}^{(0)}\exp (-i\lambda z+i\Omega
x)+b_{3}^{(0)}\exp (i\lambda ^{\ast }z-i\Omega x)]\exp (2ikz),
\end{array}
\label{perturbation}
\end{equation}
where $U_{1,2,3}^{(0)}$ stand for the amplitudes of the CW
solutions defined above, $a_{1,2,3}^{(0)}$, $b_{1,2,3}^{(0)}$ and
$\Omega $ are infinitesimal amplitudes and arbitrary real
wavenumber of the perturbation, and $\lambda$ is the eigenvalue to
be found. Substitution of these expressions into linearized
equations (\ref{evolution}) leads to a resolvability condition for
$\mu \equiv \lambda ^{2}$, in the form of a cubic equation
\begin{equation}
\mu ^{3}+\Gamma _{2}(\Omega )\mu ^{2}+\Gamma _{1}(\Omega )\mu +\Gamma
_{0}(\Omega )=0.  \label{cubic}
\end{equation}
Obviously, the stability requires that $\mu $ must be real and positive for
all $\Omega $.

We start the analysis for the case of $D=1$. Then, the
coefficients in Eq. (\ref{cubic}) are

\begin{equation}
\begin{array}{l}
\fl \Gamma _{2}=-\frac{1}{4}\Omega ^{4}-(\frac{1}{2}\alpha +2)\Omega
^{2}-2\alpha k-\frac{1}{4}\alpha ^{2}+4k-2\alpha , \\
\fl \Gamma _{1}=\frac{1}{2}\Omega ^{6}+(\alpha +1-k)\Omega
^{4}+(\frac{1}{2} \alpha ^{2}-2\alpha k+\alpha -\alpha
k^{2}-4k^{2}-4k)\Omega ^{2}\ -k\alpha
^{2}-8k\alpha -8k^{2}\alpha , \\
\fl \Gamma _{0}=-\frac{1}{4}\Omega ^{2}(-4k\alpha +\alpha \Omega
^{2}-4\alpha +\Omega ^{4})(-4k^{2}-4k+\alpha +\Omega ^{2}),
\end{array}
\label{PR-MI}
\end{equation}
for the CW solution of the PR type. For the PI-type solution, they are

\begin{equation}
\begin{array}{l}
\fl \Gamma _{2}=-\frac{1}{4}\Omega ^{4}-(\frac{1}{2}\alpha +2)\Omega
^{2}-2\alpha k-\frac{1}{4}\alpha ^{2}-4k+2\alpha , \\
\fl \Gamma _{1}=\frac{1}{2}\Omega ^{6}+(\alpha +1+k)\Omega
^{4}+(\frac{1}{2} \alpha ^{2}+6\alpha k+3\alpha -\alpha
k^{2}-4k^{2}+4k)\Omega ^{2}\ +k\alpha
^{2}-8k\alpha +8k^{2}\alpha , \\
\fl \Gamma _{0}=\frac{1}{4}\Omega ^{2}(4k\alpha +\alpha \Omega
^{2}-4\alpha +\Omega ^{4})(4k^{2}-4k-\alpha -\Omega ^{2}).
\end{array}
\label{PI-MI}
\end{equation}

\begin{figure}[tbh]
\centering \includegraphics[width=\linewidth,bb = 100 100 600 720]{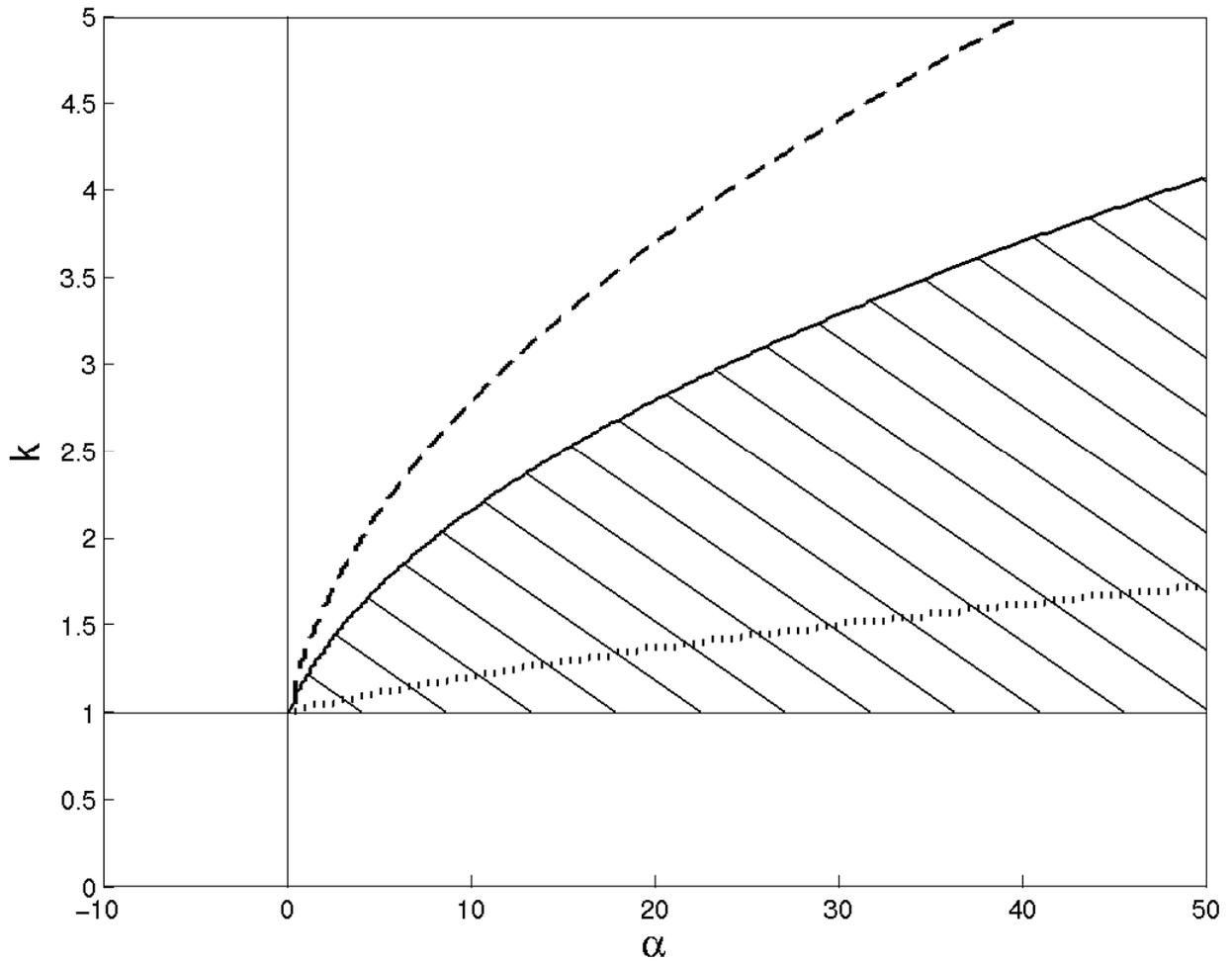} \caption{In the
shaded region in the plane of the parameters $\left( \alpha \equiv
4k+q,k\right) $, the CW solution (\protect\ref {cw-two-wave-imag})
of the PI type is stable for $D=1$ [recall that $k$ is the
wavenumber of the solution, see Eqs. (\protect\ref{general}), and
$q$ is the phase-mismatch coefficient in Eqs.
(\protect\ref{evolution})]. The stability borders of the CW
solution for $D=0.5$ and $D=10$ are shown, respectively, by the
dashed and dotted curves.} \label{mi}
\end{figure}

Equation (\ref{cubic}) was solved numerically (an analytical
solution is formally available, but it takes an intractably
complex form). The result, in the form of the stability region of
the PI solution, is displayed in Fig. \ref{mi}. Notice that the
solution is always unstable if its wavenumber belongs to the gap
of the FF subsystem, which is $k^{2}<1$.

The CW solutions of the PR-type solutions are always
modulationally unstable. However, for large values of $|\alpha |$
(actually, in the case of large phase mismatch), the instability
band of the perturbation wavenumber $\Omega $ becomes very narrow.
For example, it is $8.69<|\Omega |<8.71$ for $\alpha =-100$ and
$k=-1.2$. Therefore, the instability generated by a random
perturbation will grow very slowly in that case, and, for
practical purposes, the CW state may be regarded stable.

The above results pertain to $D=1$ in Eqs. (\ref{evolution}). With other
values of $D$, we have found that, for the PI-type CWs, the increase of $D$
results in shrinkage of the stability area, and the decrease of $D$ results
in its expansion, see the dotted and dashed curves in Fig. \ref{mi}. The
PR-type CW solution remains unstable at any $D$.

\section{Dark solitons}

\label{dark}

\subsection{Stationary solitons}

The availability of the modulationally-stable CW background
suggests a possibility of existence of stable DSs. As concerns
search for stationary symmetric ($c=0$) solutions, it is obvious
that Eqs. (\ref{model}) are invariant with respect to the
transformation $U\rightarrow U/\sqrt{D}$, $V\rightarrow V$,
$\alpha \rightarrow \alpha /D$, and $D\rightarrow 1$, which makes
it possible to set $D=1$, varying only $\alpha $ in the stationary
equations. Further, we substitute $U(x)=p(x)+ig(x)$, which yields
equations for the real functions $p,g$ and $V$,
\begin{equation}
\begin{array}{l}
p^{\prime }=g(k-1+V), \\
g^{\prime }=-p(k+1+V), \\
V^{\prime \prime }=p^{2}+g^{2}+\alpha V.
\end{array}
\label{model_real}
\end{equation}

The SH component can be sought for as $V=V_{0}+\widetilde{V}$,
where $V_{0}$ is the CW solution (\ref{cw-two-wave-imag}) of the
PI-type. Then, for the solutions supported by the PI-type CW
background, we obtain, from Eqs. (\ref {model_real}),
\begin{eqnarray}  \label{model_real_pi}
p^{\prime } &=&g\widetilde{V},~g^{\prime }=-p(2+\widetilde{V}),
\label{model_real_pi_2} \\
\widetilde{V}~^{\prime \prime } &=&p^{2}+g^{2}+\alpha (1-k)+\alpha
\widetilde{V}.  \label{model_real_pi_3}
\end{eqnarray}
Assuming, for the time being, $\left\vert \widetilde{V}\right\vert \ll 1$,
we approximate the second equation in (\ref{model_real_pi_2}) by $g^{\prime
}=-2p$ (the validity of this assumption will be considered later). Then, we
look for a fundamental DS solution in the form
\begin{equation}
\begin{array}{l}
g=\left\vert U_{0}\right\vert \tanh (\beta
x),~p=p_{0}\mathrm{sech}
^{2}(\beta x), \\
V=V_{0}+\widetilde{V}_{0}\mathrm{sech}(\beta x),
\end{array}
\label{ansatz}
\end{equation}
where $U_{0}$ and $V_{0}$ are the same as in Eqs.
(\ref{cw-two-wave-imag}) (hence the solution is matched to the CW
background at $x\rightarrow \pm \infty $), while the constants
$p_{0},\widetilde{V}_{0}$ and $\beta $ are to be found. Note that,
according to Fig. \ref{mi}, $V_{0}\equiv 1-k$ is negative in the
region of stability of the CW background, hence the expression
(\ref{ansatz}) with $\widetilde{V}_{0}>0$ assumes that $\left\vert
V(x)\right\vert $ has a minimum at $x=0$. Substitution of the
\textit{ansatz} (\ref{ansatz}) into Eqs. (\ref{model_real_pi_2})
and (\ref {model_real_pi_3}) shows that it solves the first two
equations, but not the last one. To identify an approximate
solution, we then demand the latter equations to be satisfied only
at the central point, $x=0$. This approach leads to a system of
algebraic equations, which can be readily solved to yield the
following results:
\begin{equation}
\widetilde{V}_{0}=\frac{1}{4}\alpha \left( 1-V_{0}/4\right) \left(
\sqrt{1- \frac{8V_{0}}{\alpha (1-V_{0}/4)^{2}}}-1\right) ,
\label{approx_solution_pi_1}
\end{equation}
\begin{equation}
\beta =\sqrt{\tilde{V}_{0}},~p_{0}=-\frac{1}{2}|U_{0}|\sqrt{\tilde{V}_{0}}
\label{approx_solution_pi_2}
\end{equation}
provided that $\tilde{V}_{0}\geq 0$ and $1-\left( 8V_{0}/\alpha \right)
(1-V_{0}/4)^{-2}\geq 0$. It is straightforward to check that the latter
conditions are satisfied in the region of interest ($\alpha >0,k>1$), where
the CW solutions of the PI-type are modulationally stable, see Fig. \ref{mi}.

The above approximation assumed $\widetilde{V}_{0}\ll 1$. To
examine the validity of this assumption, we note that the
expression (\ref {approx_solution_pi_1}), considered as a function
of $\alpha $, attains the maximum value $\left(
\widetilde{V}_{0}\right) _{\max }=4V_{0}/\left( V_{0}-4\right)
\equiv 4(k-1)/(3+k)$ at $\alpha =\infty $, hence the assumption is
formally justified for small $(k-1)$ (near the upper edge of the
spectral gap) if $\alpha $ (i.e., the phase mismatch) is large,
and for larger $(k-1)$ if $\alpha $ is smaller.

According to Eqs. (\ref{ansatz}), the parameters $p_{0}$ and
$\widetilde{V}_{0}$ determine the inverse DS contrasts $\sigma
_{u,v}$, which may be defined as ratios of the amplitudes at the
soliton's center to those in the CW background: $\sigma
_{u}=|p_{0}/U_{0}|$ and $\sigma _{v}=|\widetilde{V} _{0}/V_{0}|$
for the FF and SH, respectively. These parameters belong to the
interval $0\leq \sigma _{u,v}\leq 1$, the limiting cases of
$\sigma _{u,v}=0$ and $\sigma _{u,v}=1$ corresponding,
respectively, to the black soliton, and to the unmodulated CW
solution. In particular, Eqs. (\ref {approx_solution_pi_1}) and
(\ref{approx_solution_pi_2}) predict the black soliton at $\alpha
=0$.

\begin{figure}[p]
\subfigure[$k=1.2$]{\includegraphics[scale=0.5]{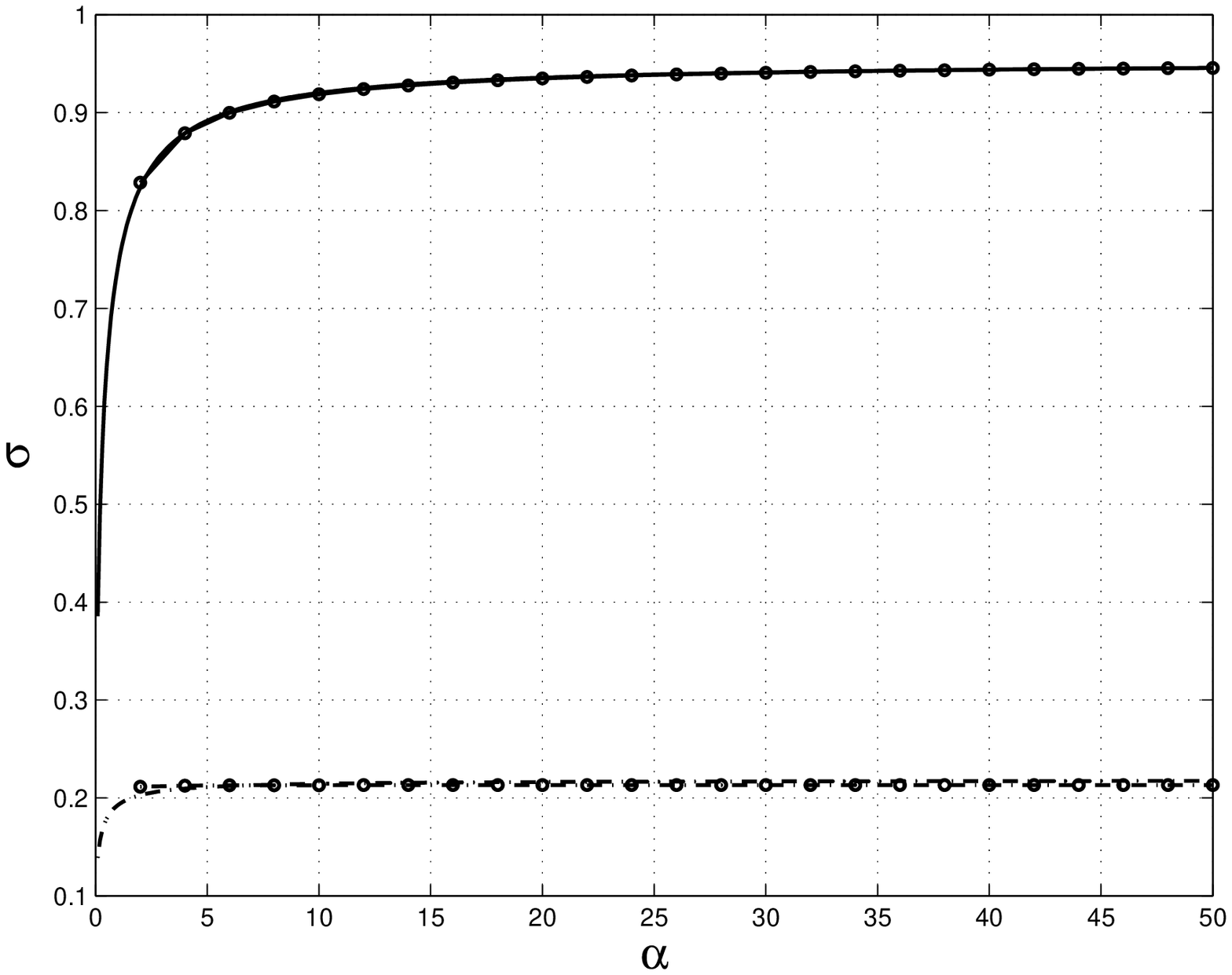}}
\subfigure[$k=2.4$]{\includegraphics[scale=0.5]{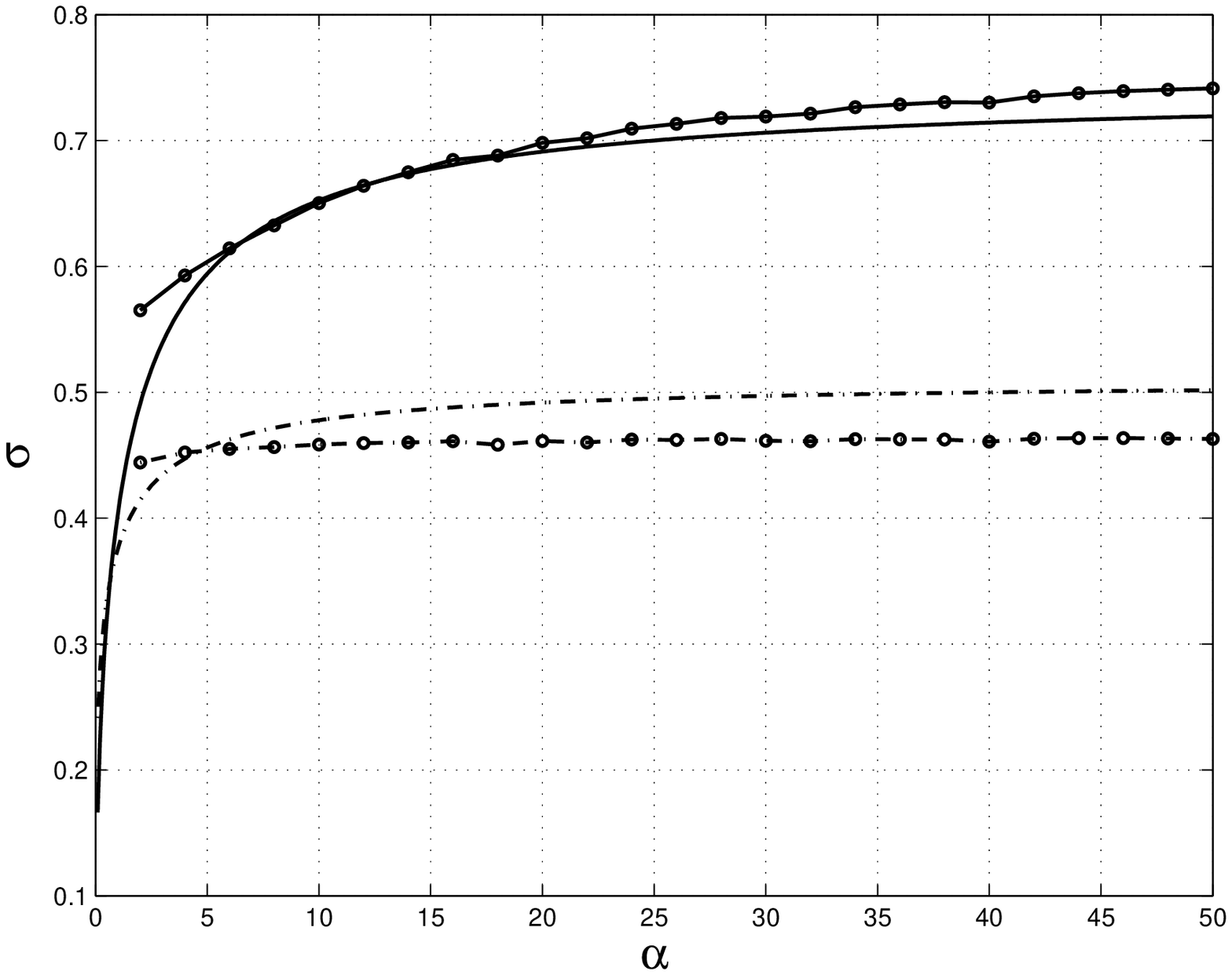}}
\caption{The inverse contrasts $\protect\sigma _{u}$ and
$\protect\sigma _{v} $ of the fundamental dark soliton supported
by the modulationally stable CW background of the PI-type. The
solid and dashed-dotted lines show, respectively, the analytical
prediction for $\protect\sigma _{v}$ and $\protect\sigma _{u}$;
curves of the same types connecting circles show numerical results
for $\protect\sigma _{v}$ and $\protect\sigma _{u}$. The panels
(a) and (b) pertain to $k=1.2$ and $k=2.4$, and in both cases
$D=1$.} \label{brightness_coeffs}
\end{figure}

The dependences of the inverse contrast on $\alpha $, as obtained
from the above approximation and from a direct numerical solution,
are shown in Fig. \ref{brightness_coeffs} (two cases shown in this
figure adequately represent the general situation). The results
demonstrate that the analytical approximation is, generally, quite
accurate, although numerically found solitons are never completely
black. With the increase of $k$, the FF component becomes
``grayer", while the SH one gets darker.

\begin{figure}[p]
\subfigure[$k=1.2$,
$\alpha=20$]{\includegraphics[scale=0.5]{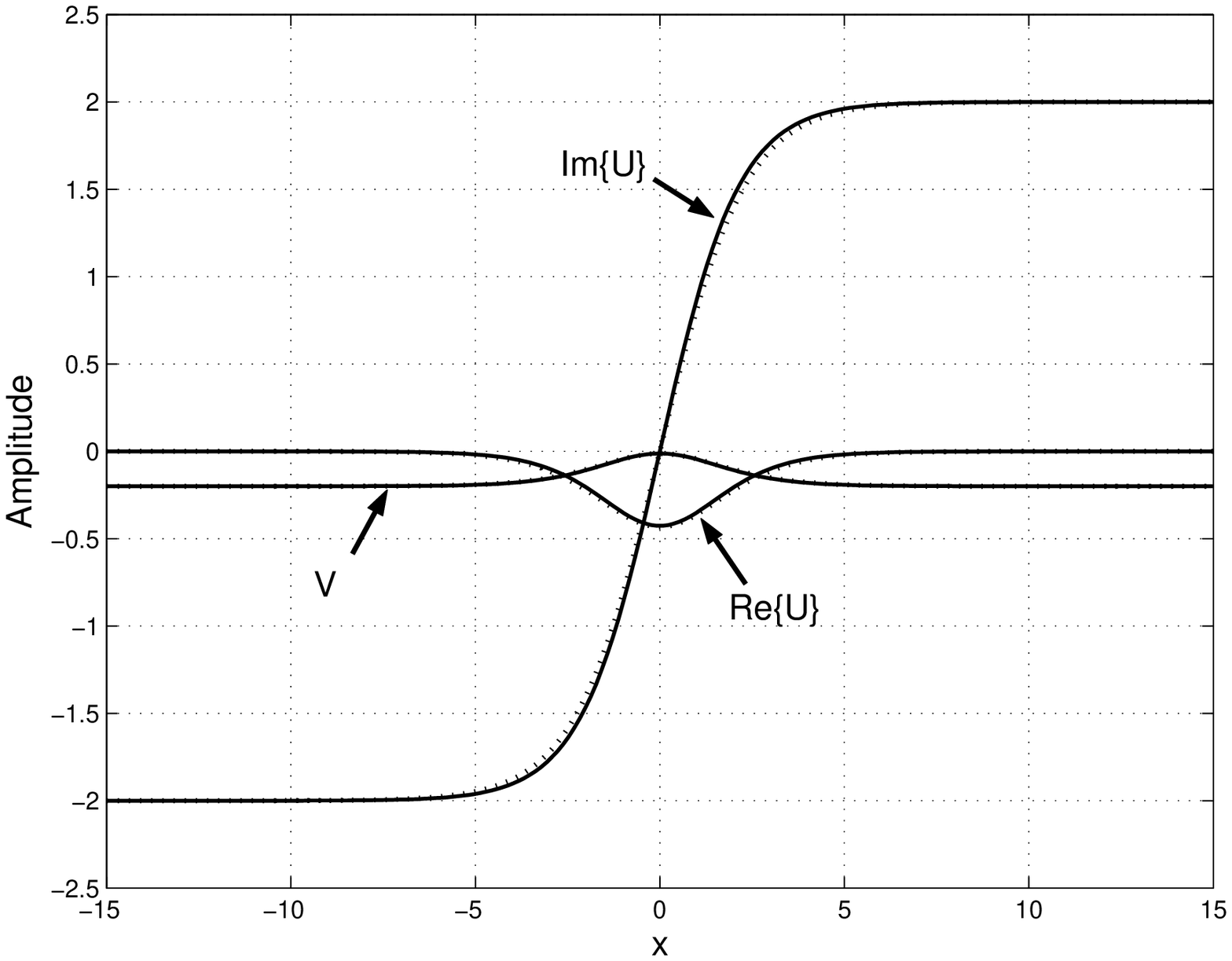}}
\subfigure[$k=2.4$,
$\alpha=40$]{\includegraphics[scale=0.5]{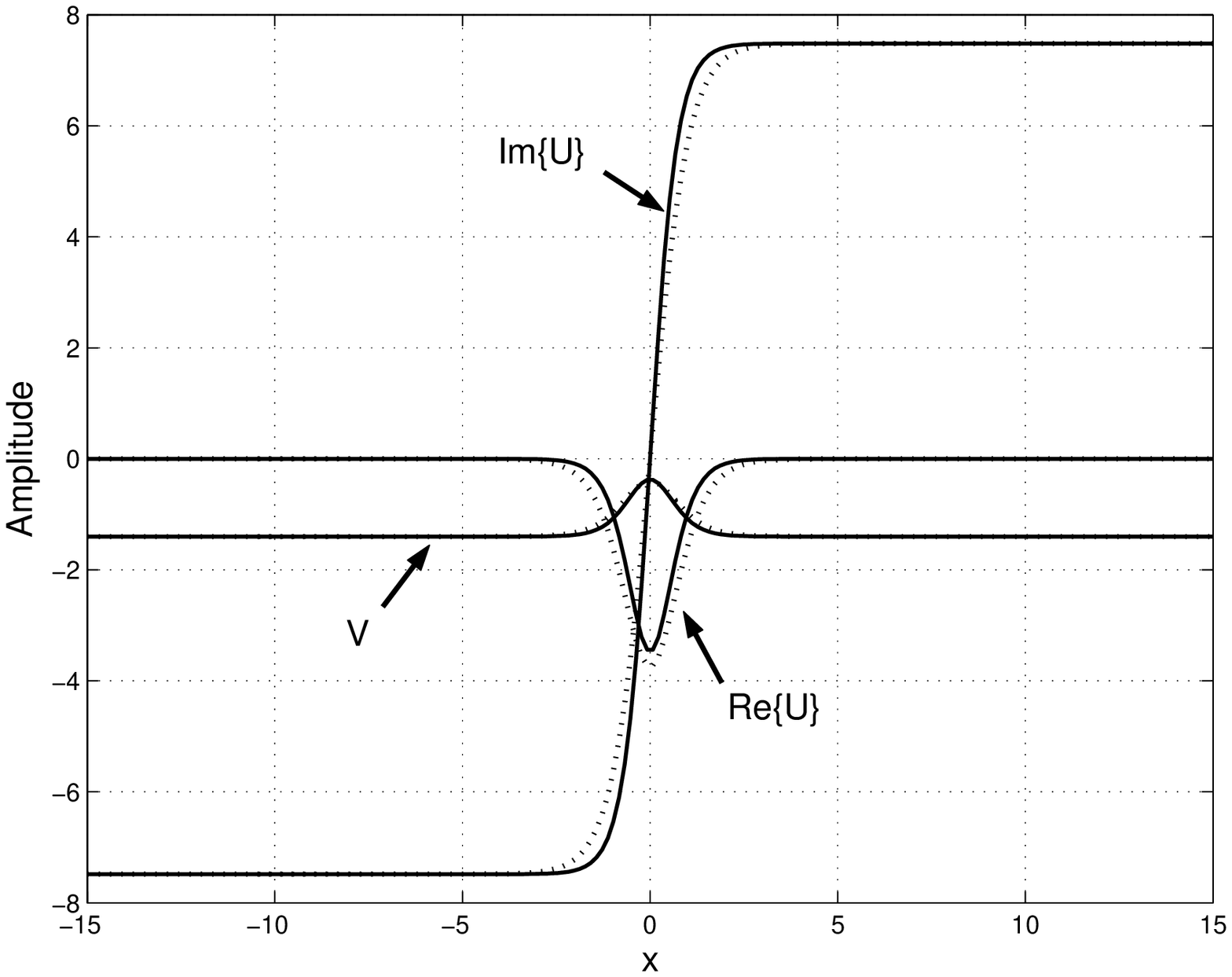}}
\caption{Typical examples of the fundamental dark soliton. The solid and
dotted lines show, respectively, the numerical solution and the analytical
approximation: (a) $k=1.2,\protect\alpha =20$; (b) $k=2.4,\protect\alpha =
40 $.}
\label{fundamental_dark}
\end{figure}

Typical examples of the direct comparison between the DS shape, as predicted
by the analytical approximation, and as generated by the numerical solution,
are displayed in Fig. \ref{fundamental_dark}. For $k=1.2$ and $\alpha =20$,
the approximation is so close to the exact solution that they are
indistinguishable. For $k=2.4$ and $\alpha =40$, the accuracy deteriorates,
but the agreement is still fairly good.

\begin{figure}[p]
\begin{center}
\includegraphics[scale=0.5]{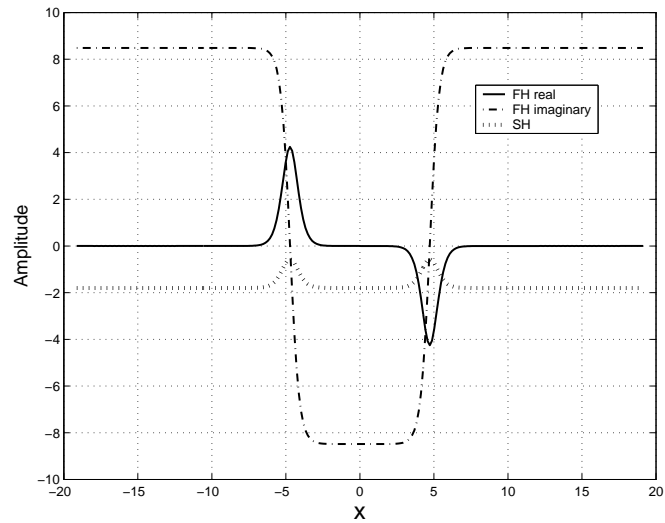}
\end{center}
\caption{A bound state of two dark solitons, found numerically at $k=2.8$
and $\protect\alpha =40$. In this figure and below, the labels FH and SH
pertain, respectively, to the fundamental and second harmonics.}
\label{twin_dark}
\end{figure}

\begin{figure}[p]
\begin{center}
\includegraphics[scale=0.5]{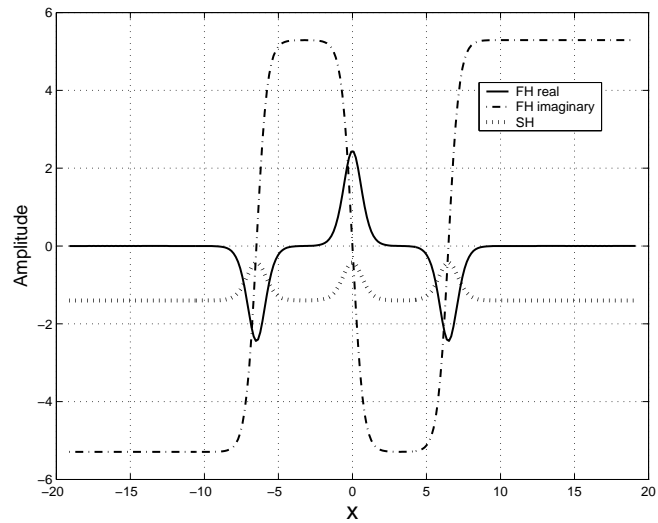}
\end{center}
\caption{A typical example of a bound state of three dark solitons, found
for $k=2.4$ and $\protect\alpha =20$.}
\label{tripple_dark}
\end{figure}

Further numerical search for DS solutions supported by the
modulationally stable CW background has revealed the existence of
bound states of the DSs, whose shape is similar to that of
unstable bound states which were reported Refs. \cite{Buryak95c}
and \cite{Conti97a} (however, the bound states may be
\emph{stable} in the present model, see below). Typical examples
of twin-hole and tri-hole DS complexes are shown in Figs.
\ref{twin_dark} and \ref{tripple_dark}.

\subsection{Tilted solitons}

The general ansatz (\ref{general}) for steady-shape solutions
admits tilted DSs with $c\neq 0$. We have found such solitons by
numerical search in the $(k,\alpha )$ plane. The above-mentioned
scaling, which allowed us to impose the normalization $D\equiv 1$
for the $c=0$ solitons, does not hold for the tilted ones.
Nevertheless, we set $D=1$ in this case too, as the additional
problem of scanning the parameter space augmented by the effective
diffraction coefficient $D$ is too complex.

\begin{figure}[p]
\begin{center}
\includegraphics[scale=0.5]{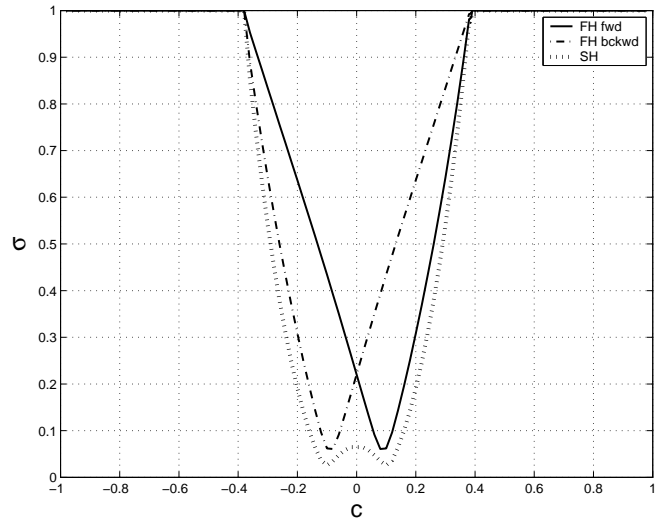}
\end{center}
\caption{The inverse contrast of three components in the tilted
(\textquotedblleft moving") fundamental dark soliton vs. its tilt
$c$ for $k=1.2$ and $\protect\alpha =20$. At points $c\equiv
c_{\mathrm{bif}}$, where all the three contrasts becomes equal to
$1$, the dark soliton disappears, merging into the CW background.}
\label{moving_sigma}
\end{figure}

For the tilted DSs, the amplitudes of the two FF components
$u_{1}$ and $u_{2}$ are no longer equal, and the SH field $u_{3}$
is not real; accordingly, one should use Eqs. (\ref{moving_model})
instead of Eqs. (\ref {model}). Numerical solution for the tilted
(\textquotedblleft moving\textquotedblright ) solitons gives rise
to the dependence of the inverse contrasts in the three components
(\textquotedblleft forward\textquotedblright\ and
\textquotedblleft backward\textquotedblright\ FF ones, and the SH
component), defined as above, on the slope (tilt) $c$. A typical
result is shown in Fig. \ref{moving_sigma}. For the forward
(backward) slope of the DS, the contrast of the forward (backward)
FF component initially increases and attains a maximum at some
tilt $c_{\max }$. For $\left\vert c\right\vert >\left\vert c_{\max
}\right\vert $, the contrasts of both FF components decrease until
the DS reaches a point $c=c_{ \mathrm{bif}}$ where $\sigma $
becomes equal to $1$, i.e., the DS disappears, bifurcating back
into the CW solution. For the forward-tilted soliton, the forward
FF component \ of the soliton is always darker than its backward
counterpart.

\begin{figure}[p]
\begin{center}
\includegraphics[scale=0.5]{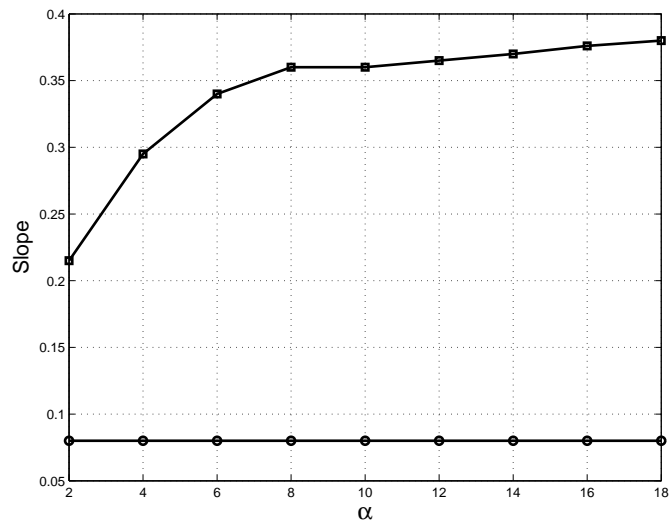}
\end{center}
\caption{The slope $c_{\mathrm{\min }}$ of the tilted dark soliton, at which
the maximum contrast of the soliton's FF component is achieved (circles),
and the slope $c_{\mathrm{bif}}$, at which the dark soliton merges into the
CW background (squares), versus the effective mismatch $\protect\alpha $,
for $k=1.2$.}
\label{sigma_of_alpha}
\end{figure}

The dependence of parameters of the tilted DS on $\alpha $, i.e., on the
phase mismatch, is shown in Fig. \ref{sigma_of_alpha}. As is seen, the slope
$c_{\max }$, at which the maximum contrast of FF occurs, is virtually
independent of $\alpha $. However, the slope $c_{\mathrm{bif}}$ at which the
tilted DS bifurcates back into the CW, i.e., the maximum slope past which
the DS does not exist, increases with $\alpha $.

\begin{figure}[p]
\begin{center}
\includegraphics[scale=0.5]{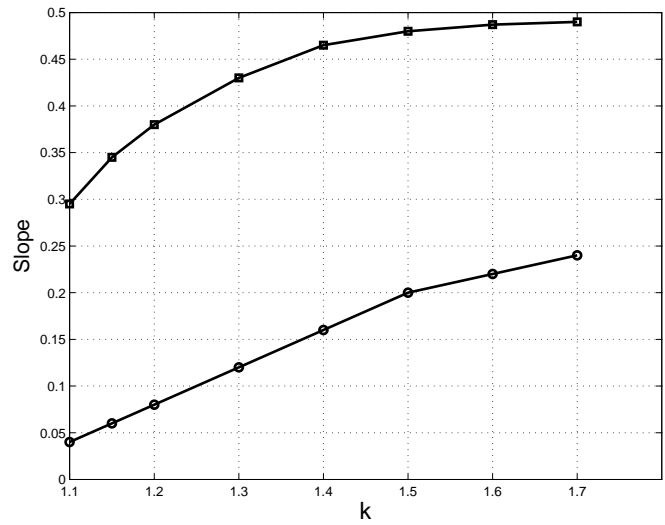}
\end{center}
\caption{The same as in Fig. \protect\ref{sigma_of_alpha}, but as a function
of $k$, for fixed $\protect\alpha =20$.}
\label{sigma_of_k}
\end{figure}

The dependence of the same parameters on $k$, i.e., as a matter of fact, on
the detuning from the gap, is displayed in Fig. \ref{sigma_of_k}. As well as
it was with the $\alpha $-dependence, the largest FF contrast is independent
of $k$, but the corresponding $c_{\max }$ grows with $k$. The bifurcation
value $c_{\mathrm{bif}}$ of the tilt is also a growing function of $k$. It
was observed too that the maximum values of the contrast remain themselves
constant with the variation of both $\alpha $ and $k$.

\begin{figure}[p]
\begin{center}
\includegraphics[scale=0.5]{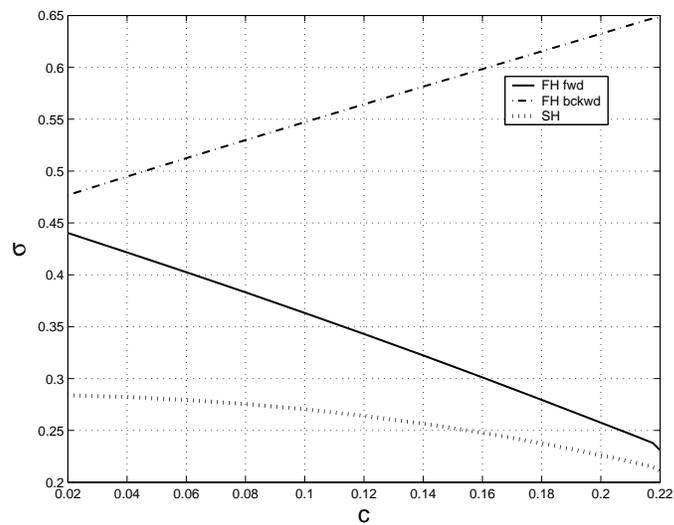}
\end{center}
\caption{The same as in Fig. \protect\ref{moving_sigma}, but for $k=2.2$.}
\label{moving_sigma_large_k}
\end{figure}

As $k$ becomes still larger than in the cases like that shown in
Fig. \ref {sigma_of_k}, viz., at $k>2.1$, the behavior of the DS
solution changes drastically. In this region, the numerical scheme
ceases to converge close to the bifurcation point, therefore the
bifurcation (merger of the DS back into the CW) cannot be
identified. A typical example of the dependence of the DS contrast
on the tilt for this situation is shown in Fig. \ref
{moving_sigma_large_k} (cf. Fig. \ref{moving_sigma}). In this
case, the numerical method does not converge for $c>0.22$.

The normalized slope $c$ relates to its physical counterpart
$c_{\mathrm{phys}}$, which is the ratio of the real coordinates,
as $c_{\mathrm{phys}}=\rho c $, where $\rho $ is the angle between
the FF wave vector and the axis $z$. Since the $\rho $ should be
small for Eqs. (\ref{evolution}) to remain valid, the physical
slopes expected to be observed in the experiment are on the order
of $10^{-3}-10^{-2}$.

\section{Stability and collisions of dark solitons}

\label{stability}

The stability of the CW background is only a necessary, but not sufficient
condition for the stability of DSs. We investigated their full stability by
direct simulations, and also through a numerical solution of the eigenvalue
problem for small perturbations. In the latter case, the perturbed solution
was taken as
\begin{equation}
\begin{array}{ccc}
u_{1}(x,z) & = & \left[ u_{1}^{(0)}(x;k)+\epsilon _{1}(x)\exp
(i\lambda z)
\right] \exp (ikz), \\
u_{2}(x,z) & = & \left[ u_{2}^{(0)}(x;k)+\epsilon _{2}(x)\exp
(i\lambda z)
\right] \exp (ikz), \\
u_{3}(x,z) & = & \left[ u_{3}^{(0)}(x;k)+\epsilon _{3}(x)\exp
(i\lambda z) \right] \exp (2ikz),
\end{array}
\label{perturbation_form}
\end{equation}
where $u_{n}^{(0)}(x;k)\ $represent an unperturbed DS solution,
$\epsilon _{n}$ are eigenmodes of infinitesimal perturbations, and
$\lambda $ is the corresponding eigenvalue, the stability
requiring all $\lambda $ to be real. The expressions
(\ref{perturbation_form}) were substituted into Eqs. (\ref
{evolution}), and the resulting equations were linearized in
$\epsilon _{n}$.

The problem significantly simplifies for the symmetric solitons
with $c=0$, when $u_{1}^{(0)}=-\left( u_{2}^{(0)}\right) ^{\ast
}$, and $u_{3}^{(0)}$ is real. In that case, we were able to find
the full spectrum of the eigenvalues for both the fundamental DSs
and their twin-hole bound states (see Fig. \ref{twin_dark}), on a
sufficiently dense grid of points in ($\alpha $,$k$) plane. As in
the previous sections, the analysis was restricted to the case of
$D=1$, as scanning through the full three-dimensional parametric
space including $D$ is a very complex technical problem. As a
result, it has been concluded that \emph{both} the fundamental and
twin-hole DSs are \emph{always stable}, provided that their CW
background of the PI type is modulationally stable.

The stability of the symmetric DSs was also verified in direct simulations
of Eqs. (\ref{evolution}). The verification has shown that they are stable
indeed, as predicted by the calculation of the perturbation eigenvalues.

Computation of the stability eigenvalues for the tilted fundamental DSs is
too hard to run it in a systematic form. However, their stability can be
easily tested in direct simulations. The simulations reveal that, as well as
their symmetric (untilted) counterparts, the tilted solitons are \emph{\
always stable} if the CW background is stable.

\begin{figure}[p]
\begin{center}
\includegraphics[scale=0.5]{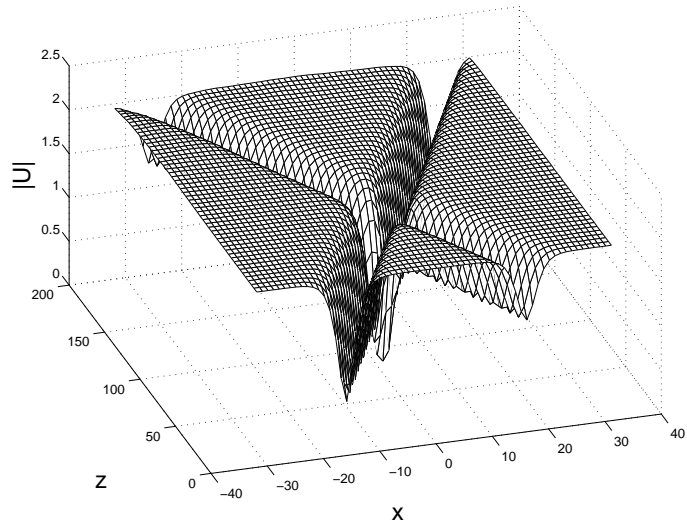}
\end{center}
\caption{A typical example of the collision between two dark solitons. The
slopes are $c=\pm 0.2$, while $k=1.2$ and $\protect\alpha =20$.}
\label{collision}
\end{figure}

\begin{figure}[p]
\begin{center}
\includegraphics[scale=0.5]{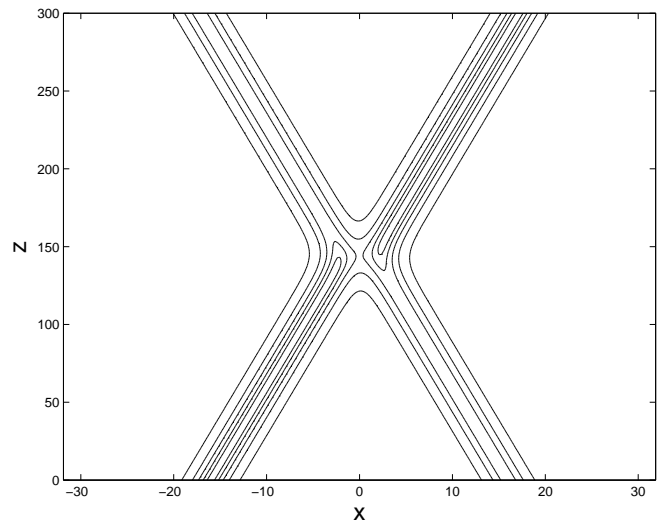}
\end{center}
\caption{Contour plot of two dark solitons collision showing the
displacement of their centers. The slopes are $c=\pm 0.1$, while $k=1.2$ and
$\protect\alpha =40$.}
\label{collision_contour}
\end{figure}

The stability of the DSs with $c\neq 0$ suggests a possibility to
consider \textquotedblleft collisions\textquotedblright\ between
them (in fact, intersection between two spatial solitons tilted in
opposite directions). The collisions were simulated too, with a
conclusion that they always look completely elastic: the DSs
re-emerge unaffected after the collision, apart from a
displacement of their centers, see an example in Fig.
\ref{collision}. The character of the interaction remains the same
for all values of $\alpha $ and $k$ (at least, for the slopes
$|c|~\geq 0.1$). The collision-induced shift of the centers is
(quite naturally) slightly larger for smaller slopes. A typical
X-junction formed by the intersection of two dark solitons is
shown in Fig. \ref{collision_contour}.

The collisions are, actually, quite similar to those observed in the
integrable NLS model describing dark solitons in a defocusing Kerr medium.
As the solitons collide, they repel each other, which is obvious from the
contour-plot rendition in Fig. \ref{collision_contour}. It is relevant to
note that interactions between DSs in the integrable self-defocusing NLS
equation are repulsive too (see, e.g., Ref. \cite{repulsion}).

\begin{figure}[p]
\begin{center}
\includegraphics[scale=0.5]{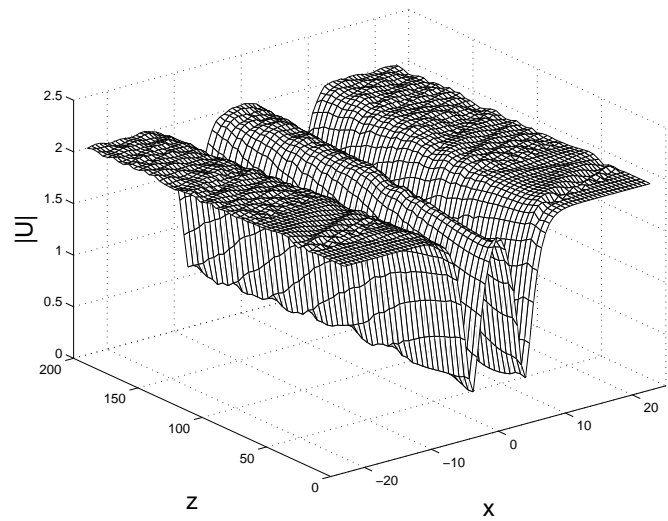}
\end{center}
\caption{A typical example of the interaction between two dark
solitons with the $c=0$ placed initially at some distance from
each other. In this case, $k=1.2$ and $\protect\alpha = 20$.}
\label{interaction}
\end{figure}

We also simulated a different type of interactions between the DS,
when two solitons with $c=0$ are originally placed at some
distance; obviously, this situation may be relevant to the
experiment, and to possible applications. It has been observed
that the solitons repel each other in this case as well. With very
small radiation loss, they start to \textquotedblleft
move\textquotedblright\ (i.e., acquire a finite slope), as is
shown in Fig. \ref{interaction}. The eventual slope observed in
this set of the simulations virtually does not depend on $k$, and
slightly increases with $\alpha $.

\section{Conclusions}

\label{conclusions}

In this work, we have considered uniform CW states and dark soliton (DSs) in
the system of three waves in a planar waveguide [with two
fundamental-frequency (FF) and one second-harmonic (SH) components] coupled
by the $\chi ^{(2)}$ interaction. The FF waves are also linearly coupled by
reflections on the Bragg grating. This model was earlier shown to generate a
great variety of bright solitons.

We have demonstrated that, on the contrary to the usual situation in
spatial-domain $\chi ^{(2)}$ models, in the present case CW states with the
phase shift $\pi /2$ between the FF and SH waves are modulationally stable
in a broad parameter region, provided that the CW wavenumber lies outside
the system's spectral gap. Stationary fundamental DSs supported by the
stable CWs were found numerically, and also by means of an \textit{ad hoc}
analytical approximation, which produced results that agree well with
numerical findings. Bound states of two and three DSs were found too. The
fundamental DSs, as well as dark bi-solitons, are stable in all the cases
when the CW background is stable; this was demonstrated by means of
calculation of the corresponding stability eigenvalues, and verified in
direct simulations.

We have also studied tilted fundamental DSs. It was found that they attain a
maximum contrast at a finite value of the slope, that virtually does not
depend on the $\chi ^{(2)}$ phase mismatch; at a maximum value of the slope
(which grows with the mismatch), the DS merges into a CW state. Interactions
between tilted solitons were systematically simulated too, with a conclusion
that the interactions are repulsive and completely elastic.

\newpage

\end{document}